\documentclass[twocolumn,showpacs,amsmath,amssymb,pra]{revtex4}

\usepackage{graphicx}
\usepackage{dcolumn}
\usepackage{bm}
\usepackage{bbm}
\usepackage{amsfonts}

\newcommand{\bra}[1]{\langle #1|}
\newcommand{\ket}[1]{|#1\rangle}
\newcommand{\braket}[2]{\langle #1|#2\rangle}

\newcommand{\tr}[2]{{\,\rm tr_{#1}}{\lbrack #2 \rbrack}\,}

\begin{document}

\title{Eigenlevel statistics of the quantum adiabatic algorithm}
\author{Marko \v Znidari\v c}
\affiliation{Department of Quantum Physics, University of Ulm, D-89069 Ulm, Germany}
\affiliation{Department of Physics, Faculty of Mathematics and Physics, University of Ljubljana, SI-1000 Ljubljana, Slovenia}

\begin{abstract}
We study the eigenlevel spectrum of quantum adiabatic algorithm for
3-satisfiability problem, focusing on single-solution instances. The properties of the ground state and the associated gap, crucial for determining the running time of the algorithm, are found to be far from the predictions of random matrix theory. The distribution of gaps between the ground and the first excited state shows an abundance of small gaps. Eigenstates from the central part of the spectrum are, on the other hand, well described by random matrix theory.  
\end{abstract}

\pacs{03.67.Lx,05.45.Mt} 

\date{\today}

\maketitle

\section{Introduction}
The question of how powerful quantum computers really are remains to be
answered. The difficulty of this question is not particular to quantum
computational complexity. Classical question of whether there exists a
polynomial algorithm for nondeterministic polynomial (NP) problems is one of
the greatest problems in mathematics. The prevailing opinion is that no such
algorithm exists. Proving this seems to be exceedingly hard. One actually has
no real idea of how to attack the problem. Due to an unintuitive character of
quantum theory similar question for quantum algorithms seems only to be
harder. Recently quantum adiabatic algorithm has been suggested for which the
initial numerical simulations showed polynomial scaling of the average running
time~\cite{Farhi:00,Farhi:01} for NP-complete problem. There are no known
classical polynomial algorithms for NP-complete problems and some plausible
arguments hint that it seems unlikely that a construction of quantum
polynomial algorithm is possible~\cite{Shor:04}. Nevertheless, even if the worst case complexity of quantum adiabatic algorithm is exponential, they might still provide a speed up for the average case performance. There were many subsequent numerical studies of the scaling of running time of adiabatic algorithm for different NP-complete problems, some indicating exponential~\cite{Smelyanskiy:01,Znidaric:05}, some polynomial dependence~\cite{Hogg:03}. While there exist analytic results for certain adiabatic algorithms ({\em e.g.} for Grover's search algorithm)~\cite{Farhi:98,Roland:02,vanDam:01,Reichardt:04}, theoretical understanding of adiabatic algorithms for NP-complete problems is still lacking. An exception is an analytical asymptotic expression for the energy gap which decreases exponentially for a particular choice of an initial Hamiltonian~\cite{Znidaric:05PRL}. In view of the conflicting numerical results and in particular due to relatively small problem sizes amenable to numerical calculation theoretical understanding is greatly desired. 

Recently random matrix theory (RMT) has been used to analyze adiabatic
algorithm~\cite{Mitchell:05,Boulatov:05}, even though it has been
noted~\cite{Boulatov:05} that it is not clear whether RMT applies to the low
energy states. If RMT description would turn out to be applicable, we could
use it to predict the behavior of adiabatic algorithm for large problem
instances. In the present paper we are going to study statistical properties
of eigenstates of adiabatic algorithm with the special emphasis on the
question whether random matrix theory is applicable. While numerical results
for small problems in Ref.~\cite{Mitchell:05} supported the usage of RMT, we
are going to show that for larger problems the behavior is quite different for
problems having a non-degenerate ground state. 

In Ref.~\cite{Mitchell:05} the failure probability of quantum adiabatic
algorithm is analyzed assuming random matrix theory spectrum, taking into
account a cascade of Landau-Zener type transitions. From the result obtained,
an exponential scaling of the running time is suggested. They also numerically
studied the distribution of ground state gaps for small 3-satisfiability
(3-SAT) problems with $n=8$ variables. The distribution obtained showed a
level repulsion which would support the usage of RMT also for the ground
state. As we will show, the distribution obtained for the small $n$ studied is
not yet an asymptotic one and the behavior for larger $n$ is very
different. On the other hand, in Ref.~\cite{Boulatov:05} the possibility of a polynomial running time is predicted, based on analysis of two RMT models, both giving essentially equivalent results. The main contribution to the failure probability in two models studied comes from the transitions to the bulk of the spectrum. 

In this paper we will study quantum adiabatic algorithm for 3-SAT problems having exactly one solution and show that while the bulk of spectrum is indeed well described by RMT the ground and the first excited state are far from RMT. In particular, it will be shown that the distribution of gaps does not show any level repulsion for sufficiently large $n$. RMT theory is therefore of limited use in describing the dynamics of standard quantum adiabatic algorithm for single-solution 3-SAT instances. As single-solution instances are thought to be the hardest, our results are important for the worst-case performance. What happens in the average case, when the number of solutions for certain values of parameters may be large, remains to be explored. 

\section{Quantum adiabatic algorithm}

We will study quantum adiabatic algorithm for 3-SAT with the standard linear interpolation between the initial Hamiltonian $H(0)$ and the final $H(1)$,
\begin{equation}
H(t)=(1-t)H(0)+t H(1).
\label{eq:Ht}
\end{equation}
The eigenstates of $H(t)$ will be denoted by $\ket{\psi_i(t)}$ with integer index $i$ denoting energy ordering, {\em e.g.}, $\ket{\psi_0(t)}$ is the ground state. The energy and time $t$ are dimensionless. The initial Hamiltonian $H(0)$ is the sum of single-qubit Hamiltonians on each qubit, 
\begin{equation}
H(0)=\sum_{i=1}^{n}{A_i\otimes \mathbbm{1}},\qquad A_i=\frac{1}{2}\begin{pmatrix}
1 & -1 \cr
-1 & 1
\end{pmatrix},
\label{eq:H0}
\end{equation}
while the final Hamiltonian $H(1)$ is a sum of $m$ three-qubit projectors, one for each clause,
\begin{equation}
H(1)=\frac{4}{\alpha} \sum_{i=1}^m{\ket{C_i}\bra{C_i}}.
\label{eq:H1}
\end{equation}
A three-qubit projector given by $C_i$ projects on the subspace of states that
violate $i$-th clause. The Hamiltonian $H(1)$ therefore simply counts the
number of clauses violated by a given computational state. Somewhat
unconventional prefactor $4/\alpha$, with $\alpha=m/n$, in $H(1)$ is chosen in
order to have time-independent trace, $\tr{}{H(t)}=N n/2$, with $N=2^n$ being
the dimension of the Hilbert space. Two relevant parameters for 3-SAT are the number of variables $n$ and the ratio of number of clauses and variables, $\alpha=m/n$. We used randomly generated 3-SAT instances having exactly one solution, {\em i.e.}, the so-called single-solution random 3-SAT. We will predominantly focus on instances with $\alpha=3$ as it has been recently numerically demonstrated~\cite{Znidaric:05} that the gap for such problems is much smaller than for those around the phase transition point for random 3-SAT~\cite{Selman}. In addition, single-solution random 3-SAT instances with small $\alpha$ also seem to be hard for classical algorithms~\cite{Znidaric:05AI}. Because previous studies focused on single-solution instances around the phase transition point we will for comparison occasionally also show the results for single-solution random 3-SAT instances with $\alpha=5$, {\em i.e.}, approximately at the location of the transition point for small $n$.
\begin{figure}[!h]
\centerline{\includegraphics[width=3.3in]{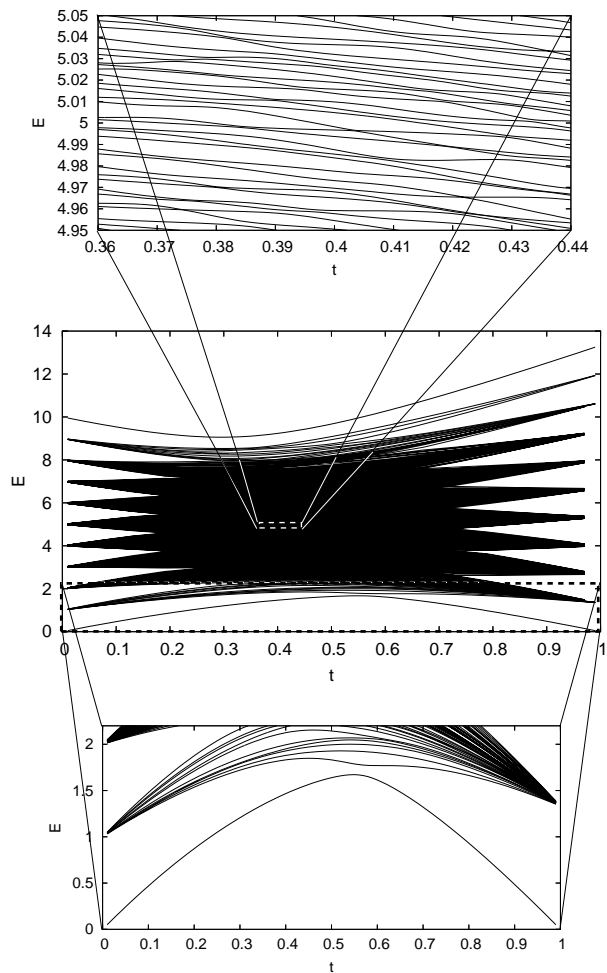}}
\caption{Spectrum for 3-SAT with $n=10$ variables and $\alpha=3$. The ground
  state gap occurring at $t_{\rm min}=0.562$ has the value $\Delta=0.111$. The
  top figure shows enlargement of a tiny part of the spectrum in the so-called
  RMT core, while the bottom one shows the lower part of the spectrum. In the top figure, showing about 20 times smaller energy scales as the bottom one, there are many avoided crossings typical for RMT level flow, while in the bottom one only one avoided crossing between the two lowest levels is dominating.}
\label{fig:test5eig}
\end{figure}
\par
To give an impression of how the eigenspectrum of $H(t)$ (\ref{eq:Ht}) looks
like, we show in Fig.~\ref{fig:test5eig} an example for $n=10$. Two
enlargements are shown, one for the central part of the spectrum and the other
one for the lowest energies. Already at first sight one can see that there is
a qualitative difference between both energy regions. While in the central
part one has many avoided crossings, typical for a level flow described by
RMT, in the lowest levels only one avoided crossing is prominent. In
Ref.~\cite{Znidaric:05} it has been numerically shown that the failure
probability is perfectly described by the Landau-Zener formula, taking into account this single avoided crossing. One would expect the eigenlevel statistics in the central part of the spectrum to be well described by RMT theory, {\em i.e.}, we will have level repulsion due to many avoided crossings while on the other hand the level repulsion is expected to be very weak for the lowest levels (if present at all).
\par
RMT is successfully used to describe spectral statistic of complex systems,
{\em e.g.}, those with chaotic classical limit~\cite{RMT}. As the parameters
of the system are changed, changing the dynamics from integrable to chaotic,
the nearest neighbor spacing distribution changes from Poissonian to Wigner's
surmise as predicted by RMT. Sometimes the perturbing parameter can be scaled
out of the system and the transition from Poisson to RMT level spacing occurs
as one goes from low to high energies in the spectrum. Such is the case, for
instance, for the hydrogen atom in a uniform magnetic field, see {\em e.g.},
Ref.~\cite{Wintgen:87}. One of Wigner's main motivations to introduce RMT has
been to describe resonances of neutron scattering on nuclei, {\em i.e.}, the
eigenspectra of nuclei. Within standard RMT the Hamiltonian has independent
matrix elements between all levels. To describe excitations of nuclei it is
actually more natural to use the so-called 2-body random matrix ensembles
having only 2-body random interactions~\cite{2body}, for a recent review, see
Ref.~\cite{Benet:03}. In particular, for 2-body random ensembles the
distribution of spacings between the ground and the first excited state is
more similar to semi-Poisson distribution $p_{\rm sP}(s)=4s\exp{(-2s)}$ than
to Wigner's surmise for Gaussian orthogonal ensemble (GOE) $p_{\rm W}(s)=\frac{\pi}{2} s \exp{(-s^2 \pi/4)}$~\cite{Flores:01}. Note that for the semi-Poisson distribution there are more small spacings than for the GOE result. One can also argue that for a typical physical system the low energy spectrum will be dominated by some quasi-excitations, {\em e.g.}, expanding the potential around the minimum one gets phonon-like excitations. Low energy spectral fluctuations as given, for instance, by the distribution of nearest neighbor spacings are therefore expected to be closer to those for integrable systems than to chaotic ones, {\em i.e.}, more Poisson-like. Furthermore, it has been recently shown~\cite{Latorre:04,Banuls:05} that the degree of entanglement of the ground state for the quantum adiabatic algorithm for exact cover problem is much smaller than the maximal possible (as it would be, for example, for random vectors).
\par
All these results suggest that the low energy properties of quantum adiabatic algorithm could significantly deviate from those given by RMT. The aim of this paper is to show that this is indeed the case. But first, let us look at the high energy part of the spectrum where we expect RMT to hold.  

\section{Bulk properties}

\begin{figure}[!h]
\centerline{\includegraphics[angle=-90,width=3.3in]{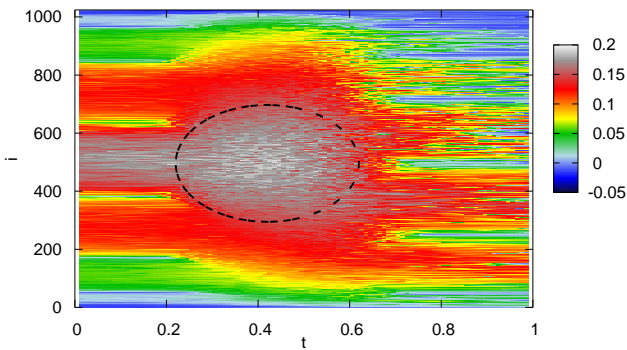}}
\caption{(Color online) Violation of the PPT criteria for eigenstates, $i=0,\ldots ,2^n-1$, at different times $t$. Color (gray) encodes the average minimal eigenvalue of $\rho^{T_A}$, [Eq.~(\ref{eq:PPT})]. Inside the black circle the average value is larger than $0.15$; we call this region the RMT core. All is for the same 3-SAT instance shown in Fig.~\ref{fig:test5eig}.}
\label{fig:test5ppt}
\end{figure}
We first test entanglement properties of eigenstates. For two qubits the positive partial transposition (PPT) is a necessary and sufficient condition for a state to be separable~\cite{Peres:96,Horodecki:96}. For $i$-th eigenstate $\ket{\psi_i(t)}$ at time $t$ the reduced density matrix $\rho_{jk}$ of the $j$-th and $k$-th qubit is obtained by tracing over all other qubits, $\rho_{jk}=\tr{\cal E}{\ket{\psi_i(t)}\bra{\psi_i(t)}}$, where ${\cal E}$ is a set of all qubits apart from $j$-th and $k$-th. We will denote the matrix obtained by partial transposition with respect to one qubit by $\rho^{T_A}_{jk}$. If the smallest eigenvalue $\lambda_{\rm min}(i,jk,t)$ of $\rho^{T_A}_{jk}$ is negative, the state $\rho_{jk}$ is entangled, otherwise, it is separable. We will use the smallest eigenvalue $\lambda_{\rm min}$ to measure 2-qubit entanglement of eigenstates. To obtain a quantity independent of two qubits $j$ and $k$ we will also average it over all pairs of qubits,
\begin{equation}
\lambda_{\rm min}(i,t)=\frac{2}{n(n-1)}\sum_{j\neq k}{\lambda_{\rm min}(i,jk,t)}.
\label{eq:PPT}
\end{equation}
The dependence of $\lambda_{\rm min}(i,t)$ on the eigenvalue index $i$ and time is shown in Fig.~\ref{fig:test5ppt}. If the eigenvectors are well described by RMT $\lambda_{\rm min}(i,t)$ should be large and positive. The reason is that for a random vector tracing over many qubits will result in a reduced density matrix that is very similar to the completely mixed one which is in turn separable. Therefore, while random vector on $n$ qubits almost certainly represents an entangled state for any bipartite cut, say $n/2+n/2$ qubits, it almost certainly does not give an entangled state when tracing over many qubits, specifically for a $2\times2$ degree of freedom reduced density matrix. Numerical simulation for random states (expansion coefficients are random Gaussian numbers) and $n=10$ qubits gives $\lambda_{\rm min}({\rm random})=0.21$, which is close to the largest values attained for eigenvectors in Fig.~\ref{fig:test5ppt}. The value $\lambda_{\rm min}({\rm random})$ saturates for large $n$, while it weakly increases with $n$ for smaller $n$. The asymptotic value is determined by the so-called induced measure on the space of density matrices~\cite{Zyczkowski:01}. Quantities other than $\lambda_{\rm min}$ can also be considered and calculated for random $2\times2$ degree of freedom matrices~\cite{Zyczkowski:98}. In Fig.~\ref{fig:test5ppt} one can observe that the largest values of $\lambda_{\rm min}(i,t)$ are obtained in a certain ``circle'' of times $t$ and eigenstates $i$. In Fig.~\ref{fig:test5ppt} we mark with a dashed curve the region where this value exceeds $0.15$. We are going to call this central portion of the spectrum a RMT core, because eigenspectrum in this region can be well described by RMT. Note that in contrast to the RMT core there is a weak 2-qubit entanglement present (negative $\lambda_{\rm min}$) in the lower part of the spectrum ({\em e.g.}, ground state).     
\par
\begin{figure}[!h]
\centerline{\includegraphics[angle=-90,width=3.3in]{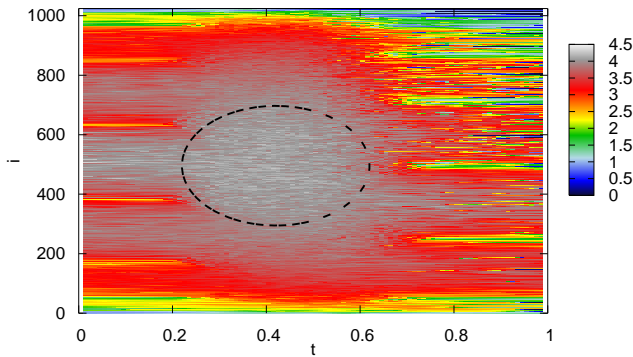}}
\caption{(Color online) The dependence of the entropy $S(i,t)$ (\ref{eq:S}) of the reduced density matrix for the first half qubits for eigenstates (index $i$) and time $t$. The entropy for maximally mixed state would be $n/2=5$. The black circle has the same location and the size as in Fig.~\ref{fig:test5ppt} and the 3-SAT instance is also the same.}
\label{fig:test5S}
\end{figure}
RMT behavior of states in the RMT core is confirmed also by studying bipartite entanglement. We divide $n$ qubits into two halves and calculate the reduced density matrix $\rho_{n/2}$ of the first $n/2$ qubits. The von Neumann entropy of this reduced density matrix then characterizes bipartite pure state entanglement. The dependence of the entropy $S(i,t)$ of the $i$-th eigenstate at time $t$,
\begin{equation}
S(i,t)=-\tr{}{\rho_{n/2} \ln_2{\rho_{n/2}}},  
\label{eq:S}
\end{equation}
is shown in Fig.~\ref{fig:test5S}. One can see the same structure as for the
PPT criteria in Fig.~\ref{fig:test5ppt}, with the highest entropy eigenstates
occurring at the same place as the highest values of $\lambda_{\rm min}$ (the
same black circle in the two figures). For low energy eigenstates we again have strong deviations from RMT, {\em i.e.}, small values of the entropy. Remember that high values of $S(i,t)$ indicate strong bipartite entanglement while large values of $\lambda_{\rm min}(i,t)$ indicate an absence of 2-qubit entanglement. 
\par
\begin{figure}[!h]
\centerline{\includegraphics[angle=-90,width=3.3in]{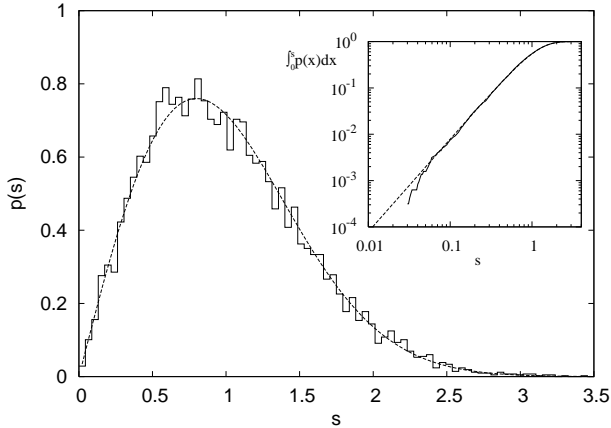}}
\caption{Level spacing distribution for states from the RMT core. $9523$ central eigenlevels of a single 3-SAT instance with $n=14$ and $\alpha=3$ are used. Dashed curve is Wigner's surmise for GOE. In the inset a cumulative distribution is show, confirming the agreement with GOE also at small spacings.}
\label{fig:rmtcore}
\end{figure}
As a final test of RMT properties of states in the RMT core we studied the nearest-neighbor level spacing statistics, the paradigmatic signature of RMT. We diagonalized the Hamiltonian at time $t=0.5$ for one 3-SAT instance with $n=14$ variables and $\alpha=3$, obtaining all $N=16384$ eigenvalues. For the central RMT core we choose to take $9523$ eigenenergies in the range $5.69 < E_i < 8.33$ (the whole spectrum lies between $1.97$ and $13.00$). Unfolding has been done by fitting a cubic polynomial to the cumulative density in the used energy interval. The level density in this region is almost constant ($\approx 2.5\cdot 10^{-4}$ in our case) and independent of a particular 3-SAT instance. The resulting level spacing distribution is shown in Fig.~\ref{fig:rmtcore} together with the Wigner's surmise $p_{\rm W}(s)=s\pi/2 \exp{(-s^2 \pi/4)}$ for the GOE ensemble. One can see nice agreement with RMT also for small spacings seen in the cumulative distribution shown in the inset of Fig.~\ref{fig:rmtcore}.  

\begin{figure}[!h]
\centerline{\includegraphics[angle=-90,width=3.3in]{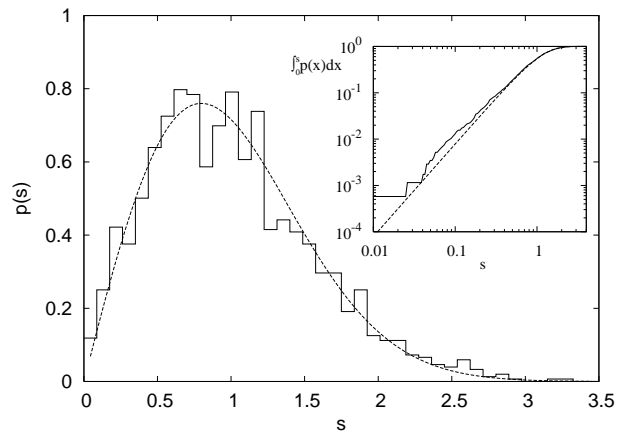}}
\caption{Level spacing distribution for low energy states ($1724$ low energy states are used, starting with the $8$th level). In the inset, one can clearly see that there are more small spacings than for the GOE ensemble.}
\label{fig:lowenergy}
\end{figure}

The situation is quite different in the low energy part of the spectrum. If using $1724$ energies in the range $2.5 < E_i < 5$ (we start with the $8$-th lowest energy), doing again cubic unfolding, we get the level spacing distribution shown in Fig.~\ref{fig:lowenergy}. While $p(s)$ might seem to be in accordance with the GOE at first sight, the behavior of the cumulative distribution, shown in the inset, reveals that there are too many small spacings as compared to the GOE result. The level repulsion is therefore weaker in the low energy region. 
\par
The properties of eigenstates in the central part of the spectrum (RMT core) are therefore well described by RMT while there are deviations for low energy states. As the ground state and the first excited state are crucial for working of quantum adiabatic algorithm we will in the next section concentrate exclusively on the properties of the ground state.

\section{Ground state gap}

A necessary condition for quantum algorithm to offer exponential advantage
over the classical is that the quantum states are sufficiently entangled,
meaning that the entanglement, as quantified, for instance, by the maximal
Schmidt number, grows exponentially with size. If this is not the case, one
could efficiently simulate quantum evolution on a classical computer~\cite{Vidal:03}. To describe the degree of entanglement we looked at the eigenvalues $\lambda_j$, $j=0,\ldots,2^{n/2}-1$, of the reduced density matrix for the first $n/2$ qubits. Square roots of this eigenvalues are Schmidt coefficients for the $n/2+n/2$ partition,
\begin{equation}
\ket{\psi}=\sum_{j=0}^{2^{n/2}-1}{\sqrt{\lambda_j} \ket{x^{\rm A}_j} \otimes \ket{x^{\rm B}_j}},
\label{eq:schmidt}
\end{equation}
where $\ket{x^{\rm A}_j}$ and $\ket{x^{\rm B}_j}$ are the corresponding eigenvectors on the first and second $n/2$ qubits, respectively. How fast the eigenvalues $\lambda_j$ decrease with $j$ will tell us the degree of entanglement and if one can use efficient methods to simulate the evolution of such states~\cite{Vidal:04}.
\begin{figure}[!h]
\centerline{\includegraphics[angle=-90,width=3.3in]{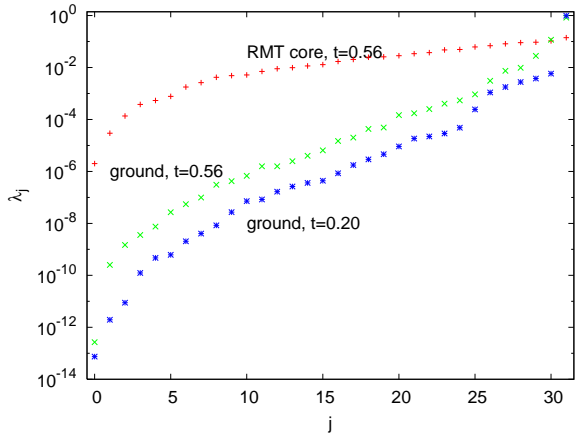}}
\caption{Eigenvalues $\lambda_j, j=0,\ldots,2^{n/2}-1$ of the reduced density matrix for the first $n/2$ qubits (\ref{eq:schmidt}). Pluses show data for a high lying eigenstate ({\em i.e.}, $512th$ eigenstate) at $t_{\rm min}=0.56$ (the location of the minimal gap), crosses are for the ground state at $t_{\rm min}=0.56$ while stars are for the ground state at $t=0.2$. All is for the same 3-SAT instance from Fig.~\ref{fig:test5eig}.}
\label{fig:test5L}
\end{figure}
In Fig.~\ref{fig:test5L} we show $\lambda_j$ for three different eigenstates: for the ground state and one high energy eigenstate from the RMT core at the location of the minimal gap and for the ground state at smaller time. One immediately notices the difference between the ground state and the state from the RMT core. In the later the eigenvalues are much larger and decrease with $j$ very slowly. On the other hand, for the ground state the eigenvalues $\lambda_j$ decrease much faster. This fast decrease of $\lambda_j$ for the ground state makes it possible to simulate ground state dynamics ({\em e.g.}, quantum adiabatic algorithm) on a much smaller space than the full $N$ dimensional Hilbert space. This has been exploited to perform numerical simulation of the quantum adiabatic algorithm for much larger $n$ than possible with the conventional methods~\cite{Banuls:05}. From Fig.~\ref{fig:test5L} we can, for instance, see that if we are content with the precision of say $10^{-3}$ one has to take $\sim 25$ eigenvectors in the case of the ``random'' state from the RMT core, while on the other hand we would need only $\sim 5$ eigenvectors for the ground state. Note that $\lambda_j$ are simply connected with the entropy $S$ [Eq.~(\ref{eq:S})] shown in Fig.~\ref{fig:test5S}. For the three cases shown in Fig.~\ref{fig:test5L} we get entropies $S(i=0,0.56)=0.90$, $S(i=0,0.20)=0.15$ and $S(i=512,0.56)=4.08$. The entropy of the ground state $S(0,t)$ attains its maximal value at the position of the minimal gap~\cite{Latorre:04}. It is significantly smaller than for the excited states from the RMT core but still grows linearly with $n$, preventing the efficient classical simulation of the quantum adiabatic algorithm for NP-complete problems~\cite{Latorre:04}.

While the entanglement of the ground state determines how efficiently we can classically simulate such an algorithm, the running time of the quantum adiabatic algorithm is predominantly determined by the minimal gap $\Delta$ between the ground and the first excited state. The necessary running time for the wanted precision at the end can be simply determined trough the Landau-Zener formula~\cite{Znidaric:05}.

In Ref.~\cite{Mitchell:05} the authors found a GOE-like distribution for the minimal gaps $\Delta$ of small 3-SAT instances with $n=8$ variables. If such behavior would persist for larger $n$ this would be advantageous because due to level repulsion we would have fewer small spacings, {\em i.e.}, the running time could be smaller. One should note that here we are talking about the distribution of ground state gaps $\Delta$ for different 3-SAT instances, {\em i.e.}, the distribution is meant over the single-solution random 3-SAT ensemble (many spectra), whereas in the previous section we looked at the distribution of spacings within a single spectrum. Here we also do not do any unfolding as there is no obvious unfolding procedure for the lowest state. 

\begin{figure}[!h]
\centerline{\includegraphics[angle=-90,width=3.3in]{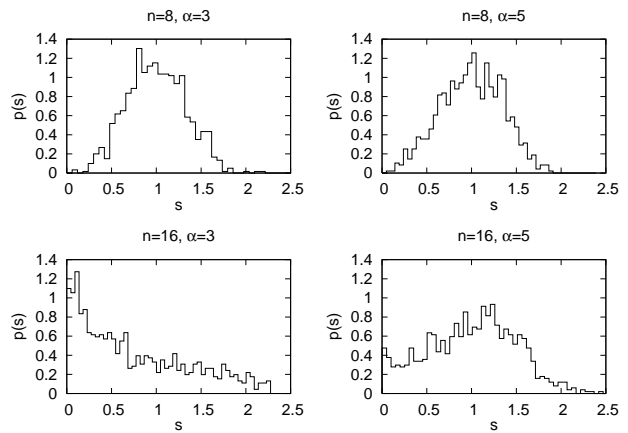}}
\caption{Histograms of the distribution of minimal gaps $\Delta$ for $1000$ single-solution 3-SAT instances for different $n$ and $\alpha$. Spacings $s$ are expressed in terms of the average spacing $\bar{\Delta}$ that can be read from Fig.~\ref{fig:gaps}. For small $n$ there are few small spacings, whereas for larger $n$ their number increases. This transition happens sooner for smaller $\alpha$.}
\label{fig:hist}
\end{figure}
In Fig.~\ref{fig:hist} we show the distribution of gaps for $1000$ random 3-SAT instances at two different $n$ and $\alpha$. The gaps are expressed in terms of the average gap, $s=\Delta/\bar{\Delta}$. We can see that for small $n$ we indeed have a sort of level repulsion. For larger $n=16$ the character is quite different though. For $\alpha=3$ the distribution is more Poisson-like with an abundance of small spacings. For $\alpha=5$ a similar behavior can be observed, but it seems that for larger $\alpha$ the change from GOE-like to Poisson-like distribution takes place at larger $n$.

\begin{figure}[!h]
\centerline{\includegraphics[angle=-90,width=3.3in]{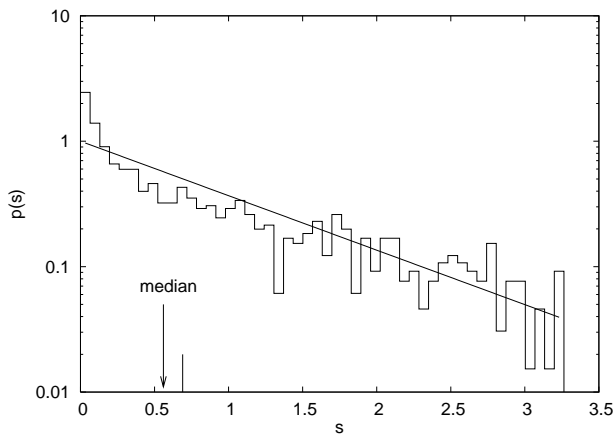}}
\caption{Level spacing distribution for 1000 single-solution random 3-SAT
  instances with $n=18$ and $\alpha=3$. The full line is the exponential curve (Poisson spacing distribution), the arrow shows the position of the median spacing, $s_{\rm med}=0.56$, and the vertical line next to the median is the value of the median for an exponential distribution. All spacings are expressed in terms of the average spacing, $s=\Delta/\bar{\Delta}$. No RMT-like level repulsion is present.}
\label{fig:hist18}
\end{figure}

The histogram for the largest case of $n=18$, $\alpha=3$, we generated is shown in Fig.~\ref{fig:hist18}. We can see that the distribution is close to Poissonian with two important differences. There are more small spacings and more large spacings than one would expect for an exponential distribution. Of course, the probability to have zero spacing is zero, so for very small spacings $p(s)$ goes towards zero, {\em e.g.}, for the case in Fig.~\ref{fig:hist18} cumulative distribution (not shown) grows as $\sim s^{1.5}$ for very small spacings of order $s \approx 10^{-3}-10^{-2}$.

\begin{figure}[!h]
\centerline{\includegraphics[angle=-90,width=3.3in]{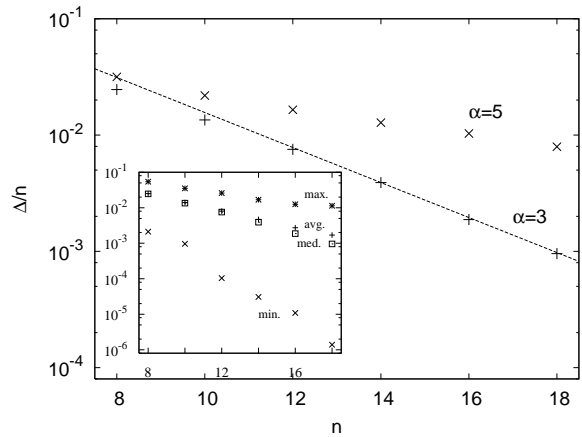}}
\caption{The dependence of the median gap $\Delta/n$ on $n$ for $\alpha=3$ and $\alpha=5$. Exponential dependence can be seen for $\alpha=3$ while for $\alpha=5$ the large $n$ dependence is hard to infer (fitting a power law gives $\Delta\sim n^{-0.6}$). Dashed line is $\Delta/n=\frac{1}{2\sqrt{N}}$. In the inset we show for $\alpha=3$ in addition to the median also the average, minimal, and maximal gap, all out of $1000$ instances.}
\label{fig:gaps}
\end{figure}

In Fig.~\ref{fig:gaps} we show in the main plot the median spacing for
$\alpha=5$ and $\alpha=3$. Instead of showing $\Delta$ directly we divided it
with $n$ because the trace of $H(t) \propto n$ and so we expect that the
eigenvalues themselves will grow proportionally to $n$. We can see that the
dependence of $\Delta/n$ on $n$ is exponential for $\alpha=3$, while the
asymptotic behavior for $\alpha=5$ has possibly not yet been reached. Such
exponential decrease of the gap, suggesting exponential running time, has
already been numerically found in Ref.~\cite{Znidaric:05}. Interestingly, the asymptotic decay rate for $\alpha=3$ agrees with $2\Delta/n \asymp 1/\sqrt{N}$ (shown with dashed line), which is the same as the analytical asymptotic result obtained in Ref.~\cite{Znidaric:05PRL} for the initial Hamiltonian being a projector to the ground state. It might be that the worst-case performance ({\em i.e.}, for small $\alpha$) is $\Delta \sim 1/\sqrt{N}$ regardless of the choice of $H_0$. Still, this issue needs to be explored in more detail. In the inset to Fig.~\ref{fig:gaps} we show for $\alpha=3$ also the minimal, maximal, and the average $\Delta/n$. While all seem to have exponential dependence on $n$, their decay rate is different. This could hint that by increasing $n$ one gets increasingly more small and more large spacings. The same conclusion has been reached from the distribution of gaps in Fig.~\ref{fig:hist18} where there is also a difference between the median and the average $\Delta/n$. The distribution of gaps therefore seems to change character with $n$ and one cannot claim that the distribution found for $n=18$ (Fig.~\ref{fig:hist18}) is already the final asymptotic distribution. In fact, there even might not exist any stationary asymptotic distribution. Whether these multiple scales of the gap distribution can be remedied by some unfolding procedure is not clear. In any case the distribution of gaps is far from RMT prediction and, unfavorably, we have many small spacings that will necessitate large running times of quantum adiabatic algorithm.  

\section{Probability flow}

So far we have identified two regions in the spectrum of quantum adiabatic
algorithm for 3-SAT. The bulk properties in the RMT core are described by RMT
while the ground state properties and the ground state gap $\Delta$ are far
from RMT predictions. In the present section we are going to explore how
important the RMT core states are for the success of adiabatic algorithm.

Let us denote the solution state by $\ket{\phi_{\rm sol}}$ ({\em i.e.}, ground state at $t=1$). At the beginning of the algorithm, at time $t=0$, the solution has approximately equal small overlap with all eigenstates, $|\braket{\phi_{\rm sol}}{\psi_i(0)}|^2 \sim 1/N$. The solution probability is therefore distributed over all eigenstates. During the evolution this probability gradually gets ``concentrated'' in the ground state, so that after we have passed the minimal gap $\Delta$ at $t_{\rm min}$ we have $|\braket{\phi_{\rm sol}}{\psi_0(t>t_{\rm min}}|^2 \sim 1$ while $|\braket{\phi_{\rm sol}}{\psi_{i > 0}(t>t_{\rm min}}|^2 \sim 0$. With time the probability therefore ``flows'' towards the ground state. So even though the RMT core occupies high energies it could be important for the adiabatic algorithm because the solution probability for small times is found also in these high energy eigenstates. To check how much the RMT core states participate in this probability flow we have calculated the total probability at time $t$ to find the solution $\ket{\phi_{\rm sol}}$ in the eigenstates higher than $i$-th,
\begin{equation}
p(i,t)=\sum_{j=i}^{N-1}{|\braket{\phi_{\rm sol}}{\psi_j(t)}|^2}.
\label{eq:p}
\end{equation}
Due to the normalization we of course have $p(0,t)=1$, while at the end of the algorithm all probability is in the ground state, $p(0,1)=1,p(i>0,1)=0$, due to our definition of the final ground state being the solution. Note that $p(i,t)$ gives the failure probability if we stop the algorithm at time $t$ and do not extend it to final $t=1$. The failure probability at the end, at $t=1$, provided we {\em e.g.}, remove all levels higher than $i$-th from time $t$ onwards is not necessarily equal to $p(i,t)$. It is probably correlated with $p(i,t)$ but with details depending on how we ``remove'' the levels at time $t$.    
\begin{figure}[!h]
\centerline{\includegraphics[angle=-90,width=3.3in]{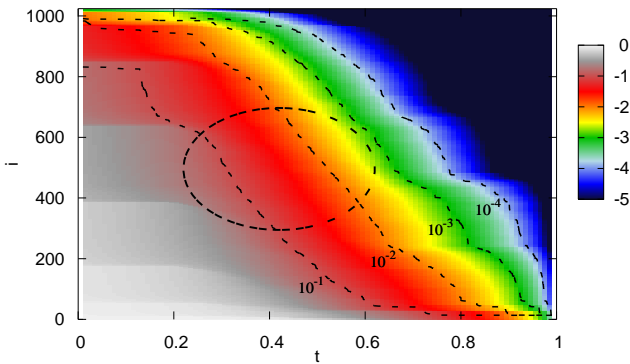}}
\caption{(Color online) Logarithm (base $10$) of the probability $p(i,t)$ [Eq.~(\ref{eq:p})] to find the solution in eigenstates higher than $i$-th. Four black curves are isolines at $10^{-1}$, $10^{-2}$, $10^{-3}$, and $10^{-4}$ probability. The black ellipse has the same location and the size as in Fig.~\ref{fig:test5ppt} and the 3-SAT instance is also the same.}
\label{fig:test5F}
\end{figure}
In Fig.~\ref{fig:test5F} we show the dependence of $p(i,t)$ for the same test 3-SAT instance with $n=10$ variables used before. We can see that the RMT core states are of limited importance. For instance, if we are satisfied with the error probability of $10\%$ ({\em i.e.}, the probability to find the solution of $0.9$), for times larger than $t\approx 0.4$ the states higher than $i\approx 300$ (location of the isoline $10^{-1}$) are not important as long as the probability flow for levels with $i<300$ stays the same. Fig.~\ref{fig:test5F} nicely illustrates that as time progresses the high energy states become less and less important. This happens already before the actual minimal gap is reached at $t_{\rm min}\approx 0.56$. One consequence of this is that the adiabatic algorithm is expected to be more insensitive to the coupling of high energy eigenstates to the environment. For some results regarding the stability of quantum adiabatic algorithms to perturbations, see~\cite{Childs:02,Roland:05,Aberg:05}.

\section{Conclusion}

We have shown that the properties of the ground state of the adiabatic algorithm for single-solution 3-SAT intances are very different from those of random vectors occurring in RMT. The distribution of gaps between the ground state and the first excited state for random 3-SAT problems with one solution shows a transition from GOE-like to Poisson-like distribution with increasing problem size. What is more, the distribution obtained for $n=18$ does not yet seem to be the asymptotic one. The ground state is also relatively weakly entangled as compared to RMT predictions. On the other hand, the central bulk portion of the spectrum is well described by RMT but has a limited influence on the flow of probability to the ground state. Therefore, RMT seems to be of limited use in describing standard quantum adiabatic algorithm for single-solution 3-SAT instances. 

\section*{Acknowledgments}
Financial support by the Alexander von Humboldt Foundation is gratefully acknowledged.

\end{document}